\documentclass[%
 reprint,
superscriptaddress,
 amsmath,amssymb,
 aps,
prstab,
]{revtex4-2}

\usepackage{graphicx}
\usepackage{dcolumn}
\usepackage{bm}
\usepackage{xcolor} 
\usepackage{multirow,ragged2e}

\begin{document}


\title{Theories derived from Haissinski equation and their applications to electron storage rings}

\author{Demin Zhou}
\email{dmzhou@post.kek.jp}
\affiliation{%
 KEK, 1-1 Oho, Tsukuba 305-0801, Japan 
}%
\affiliation{
 School of Accelerator Science, The Graduate University for Advanced Studies, SOKENDAI, Shonan Village, Hayama, Kanagawa 240-0193 Japan
}%

\author{Takuya Ishibashi}
\affiliation{%
 KEK, 1-1 Oho, Tsukuba 305-0801, Japan 
}%
\affiliation{
 School of Accelerator Science, The Graduate University for Advanced Studies, SOKENDAI, Shonan Village, Hayama, Kanagawa 240-0193 Japan
}%

\author{Gaku Mitsuka}
\affiliation{%
 KEK, 1-1 Oho, Tsukuba 305-0801, Japan 
}%
\affiliation{
 School of Accelerator Science, The Graduate University for Advanced Studies, SOKENDAI, Shonan Village, Hayama, Kanagawa 240-0193 Japan
}%

\author{Makoto Tobiyama}
\affiliation{%
 KEK, 1-1 Oho, Tsukuba 305-0801, Japan 
}%
\affiliation{
 School of Accelerator Science, The Graduate University for Advanced Studies, SOKENDAI, Shonan Village, Hayama, Kanagawa 240-0193 Japan
}%

\author{Karl Bane}
\affiliation{%
 SLAC National Accelerator Laboratory, Stanford University, Menlo Park, California 94025, USA 
}%

\author{Linhao Zhang}
\affiliation{%
 University of Science and Technology of China, No. 443, Huangshan Road, Hefei, Anhui, 230027, China 
}%

\date{\today}

\begin{abstract}
As a stationary solution of the Vlasov-Fokker-Planck equation, the Haissinski equation predicts the equilibrium line density of a bunch that circulates in a storage ring for a given wake function. This paper shows that some equations regarding the centroid shift of the bunch, the peak position of the bunch profile, bunch length, and extraction of impedance from the bunch profile can be derived from the Haissinski equation in a self-consistent manner. In particular, a generalized quadratic equation for potential-well bunch lengthening is obtained to accommodate any absolute impedance model, expanding upon Zotter's cubic equation, which is primarily applicable to inductive impedance. The equations derived in this paper are tested using computed impedance models for some electron storage rings, showing machine-dependent properties of impedance effects. We conclude that these equations can be employed in electron storage rings to effectively bridge the gap between impedance computations and beam-based measurements.
\end{abstract}

\maketitle


\section{Introduction}
Modern particle accelerators are designed to deliver high-intensity and high-brightness charged beams for experiments in a wide range of fields, from high-energy particle physics to material science. Interactions between charged beams and their surroundings, mediated by beam-induced electromagnetic fields and commonly described by the concepts of wakefield and its Fourier transform impedance~\cite{zotter1998impedances, chao1993physics}, play a crucial role in limiting beam quality. Throughout the entire life cycle of an accelerator project,  it is imperative to thoroughly investigate the collective effects driven by impedance. Typically, impedance budgets are created during the design phase to predict impedance-driven beam phenomena reliably. During machine commissioning, beam-based measurements are performed to validate predictions from previous simulations or to identify any impedance sources that may have been overlooked.

In recent decades, many theories and simulation tools have been developed to facilitate connections between impedance calculations and beam measurements~\cite{rumolo2016beam, smaluk2018impedance, salvant2019building}. For such comparisons, one concern is that the quantities extracted from the calculations and beam measurements may not be identical; proper translations between the two are sometimes required, though not always. For example, Zotter's cubic equation~\cite{zotter1981potential} has been widely used to extrapolate the longitudinal effective impedance~\cite{chao1993physics} from bunch length measurements in electron storage rings and then compare with the broad-band impedance models constructed beforehand. For such comparisons, a comprehensive review is given in~\cite{smaluk2018impedance}. However, Zotter's equation is not self-consistent and is valid only when a pure inductance (i.e. purely imaginary impedance) can approximate the longitudinal total impedance of the ring. For a review of Zotter's equation and its validity, see~\cite{zhou2023potential}. When applied to a real machine, the equation itself might introduce uncertainties and contribute to the discrepancy between simulations and experiments, in addition to other factors reviewed in~\cite{smaluk2018impedance, salvant2019building}. On the other hand, the Haissinski equation~\cite{haissinski1973exact} is a stationary self-consistent solution of the Vlasov-Fokker-Planck (VFP) equation~\cite{frank2020linear} with an absolute impedance model below the threshold current of microwave instability (MWI, see Sec.~2.4.10 of~\cite{chao2023handbook} for further details). In principle, it is more appropriate to compare the solutions of the Haissinski equation below the MWI threshold current or the VFP equation at higher currents with experimental results when studying the longitudinal impedance effects. However, certain efforts must be made to solve these nonlinear equations (for example, see~\cite{bane2010threshold, warnock2018numerical} for numerical techniques to solve these equations). To reduce such efforts, this paper presents several handy equations derived from the Haissinski equation to facilitate the calculation-experiment connections with their applications to real machines.

The paper is organized as follows. After a brief review of the Haissinski equation, in Sec.~\ref{sec:theories} we derive several equations which describe the current-denpendences of center of mass, peak position of the longitudinal profile, and bunch length. We show how these frequently measured quantities in electron storage rings are correlated with impedances in a self-consistent manner. The inverse problem of the Haissinski equation, that is, extracting the frequency-dependent impedance from a longitudinal bunch profile, is also discussed in this section. For specific impedances established in the literature, in Sec.~\ref{sec:ExplicitExpressions} we derive explicit expressions for center of mass and bunch length with a Gaussian approximation of the bunch profile. We show that Zotter's cubic equation~\cite{zotter1981potential} is a special case of our generalized ``quadratic'' equation for potential-well bunch lengthening. In Sec.~\ref{sec:applications}, the theories are tested with a few real machines, where impedance modeling and impedance effects have been intensively investigated. Finally, we summarize our findings in Sec.~\ref{sec:summary}.

\section{\label{sec:theories}Theories derived from Haissinski equation}

\subsubsection{Haissinski equation}

Individual electrons oscillate around a fixed point in the longitudinal phase space in electron storage rings. The RF system creates a potential well that confines the beam in a bucket. When the oscillations are of small amplitude, the motion of the electrons is linear. In the presence of radiation damping and quantum excitation, the beam reaches a stable equilibrium distribution in the longitudinal phase space, which is Gaussian~\cite{chao2020lectures}. The longitudinal wakefields deform the potential well created by the RF system, which, in turn, affects the wake force experienced by the beam. Consequently, it is necessary to solve the system for a self-consistent stationary distribution.

For electron storage rings, the high-intensity circulating beam can be modeled as a continuous distribution $\psi$, with its evolution governed by the VFP equation. Specifically, when considering the synchrotron motion, the VFP equation is~\cite{jowett1987introductory}
\begin{equation}
    \frac{\partial\psi}{\partial s}
    +\frac{d z}{d s} \frac{\partial \psi}{\partial z}
    +\frac{d \delta}{d s} \frac{\partial \psi}{\partial \delta}
    =\frac{2}{ct_d} \frac{\partial}{\partial \delta}
    \left[ \delta\psi + \sigma_\delta^2 \frac{\partial \psi}{\partial \delta} \right].
    \label{eq:vfp_equation}
\end{equation}
Here, $s$ is the arc length on the beam's closed orbit, $z=s-s_0=s-ct$ is the longitudinal displacement from the synchronous particle at $s=s_0=ct$, $\delta=(P-P_0)/P_0$ is the momentum deviation, $\sigma_\delta$ is the momentum spread, and $t_d$ is the longitudinal damping time. The equations of motion, including wakefields, are
\begin{equation}
    \frac{dz}{ds}=-\eta \delta, \quad \frac{d\delta}{ds}=
    \frac{\omega_s^2}{\eta c^2}z - F(z,s),
    \label{eq:Equation_of_motion_delta}
\end{equation}
with $\eta$ the slip factor and $\omega_s$ the synchrotron frequency. The wakefield term $F(z,s)$ is calculated from the convolution of charge density and longitudinal (point charge) wake function
\begin{equation}
    F(z,s)=I_n \int_{-\infty}^\infty W_\parallel(z-z')\lambda(z',s) dz',
    \label{eq:Wake_force}
\end{equation}
with the scaling factor $I_n=Ne^2/(EC_0)$ and the line density distribution $\lambda(z,s)=\int_{-\infty}^\infty \psi(z,\delta,s)d\delta$. Here, $N$ is the bunch population, $C_0$ is the circumference of the storage ring and $E$ is the energy of the reference particle. The wake function $W_\parallel(z)$ has various forms in the literature depending on the conventions chosen. In this paper, we follow the conventions of~\cite{chao1993physics, chao2023handbook}. The wake function and the corresponding impedance are the Fourier transforms of each other: $W_\parallel(z)=\frac{c}{2\pi}\int_{-\infty}^\infty Z_\parallel(k) e^{ikz} dk$ and $Z_\parallel(k)=\frac{1}{c}\int_{-\infty}^\infty W_\parallel(z) e^{-ikz} dz$.

With synchrotron radiation effects but without wakefields, the beam in an electron storage ring is bunched with an equilibrium bunch length $\sigma_{z0}$ given by $\sigma_{z0}=-c\eta\sigma_\delta/\omega_s$ (note that $\omega_s<0$ when $\eta>0$~\cite{wolski2014beam}). The VFP equation has an $s$-independent stationary solution $\psi_{ss}(z,\delta)$ in the form of $\psi_{ss}(z,\delta)=\hat{\psi}_0(\delta) \lambda_h(z)$ below the microwave instability threshold. The momentum distribution $\hat{\psi}_0(\delta)$ is Gaussian with rms spread $\sigma_\delta$ and the spatial distribution satisfies
\begin{equation}
    \frac{d\lambda_h(z)}{dz}+
    \left[ \frac{z}{\sigma_{z0}^2}-\frac{1}{\eta\sigma_\delta^2} F_h(z) \right]
    \lambda_h(z)
    =0.
    \label{eq:de_Haissinski}
\end{equation}
Here $F_h(z)$ is from Eq.~(\ref{eq:Wake_force}) with $\lambda(z,s)$ replaced by the equilibrium distribution $\lambda_h(z)$. The solution of Eq.~(\ref{eq:de_Haissinski}) is the so-called Haissinski equation~\cite{haissinski1973exact}
\begin{equation}
    \lambda_h(z)=
    A e^{-\frac{z^2}{2\sigma_{z0}^2}-\frac{I}{\sigma_{z0}}\int_z^{\infty}dz'\mathcal{W}_\parallel(z')},
    \label{eq:Haissinski_equation_explicit}
\end{equation}
with $A$ a normalization factor to satisfy $\int_{-\infty}^\infty \lambda_h(z)dz=1$, and $\mathcal{W}_\parallel(z)$ the bunch wake potential defined by
\begin{equation}
    \mathcal{W}_\parallel(z)=\int_{-\infty}^\infty W_\parallel(z-z')\lambda_h(z') dz',
    \label{eq:wake-potential}
\end{equation}
and the new scaling parameter $I=I_n\sigma_{z0}/(\eta\sigma_\delta^2)$. Using the synchrotron tune defined by $\nu_s=\omega_s/\omega_0$ with $\omega_0=c/C_0$ the revolution frequency, there is $I=-Ne^2/(2\pi\nu_sE\sigma_\delta)$~\cite{oide1990long, warnock2018numerical}. Equation~\ref{eq:Haissinski_equation_explicit} can be rewritten as
\begin{equation}
    \lambda_h(z)=
    A e^{-V(z)},
    \label{eq:Haissinski_equation_explicit-2}
\end{equation}
with $V(z)$ recognized as a potential well~\cite{cai2011linear}
\begin{equation}
    V(z)=\frac{z^2}{2\sigma_{z0}^2}+\frac{I}{\sigma_{z0}}\int_z^{\infty}dz'\mathcal{W}_\parallel(z').
\end{equation}

The stability of the Haissinski equation is beyond the scope of this paper, and the reader is referred to~\cite{cai2011linear} and the references therein.

\subsubsection{Centroid shift}
Integrating over $z$ on both sides of Eq.~(\ref{eq:de_Haissinski}) and recognizing that the center of mass of the bunch is $z_c=\int_{-\infty}^\infty z\lambda_h(z) dz$,
we obtain
\begin{equation}
    z_c(I)=I\sigma_{z0}\kappa_\parallel,
    \label{eq:centroid}
\end{equation}
with the well-known loss factor $\kappa_\parallel$ given by
\begin{equation}
    \kappa_\parallel(I)=\int_{-\infty}^\infty dz \lambda_h(z)\mathcal{W}_\parallel(z).
    \label{eq:LossFactor}
\end{equation}
Equation~(\ref{eq:centroid}) shows that the centroid shift of the bunch is exactly proportional to the loss factor scaled by $I$. Measurement of one of the two quantities will automatically produce the other. In particular, the centroid shift rate at $I=0$ is given by
\begin{equation}
    m_1\equiv \left. \frac{dz_c}{dI}\right |_{I=0}=\sigma_{z0}\kappa_{\parallel 0}.
\end{equation}
The quantity $\kappa_{\parallel 0}\equiv\kappa_{\parallel}(0)$ is the loss factor for the nominal bunch length $\sigma_{z0}$ at zero current and can be computed without difficulty when the impedance model is constructed. When validating an impedance budget that gives $\kappa_{\parallel 0}$(theory) for a storage ring, the process involves measuring $\kappa_\parallel$ as a function of $I$ and then extrapolating its slope at $I=0$ to determine $m_1$, thus obtaining $\kappa_{\parallel 0}$(measurement). The difference between $\kappa_{\parallel 0}$(theory) and $\kappa_{\parallel 0}$(measurement) can quantify the gap between the theoretical model and the actual impedance of a machine in the real part.

In terms of impedance, there is
\begin{equation}
    \kappa_\parallel=\frac{c}{\pi} \int_0^\infty \text{Re}[Z_\parallel(k)] h(k) dk
\end{equation}
with spectral power density $h(k)=\tilde{\lambda}_h(k)\tilde{\lambda}_h^*(k)$ where $\tilde{\lambda}_h(k)$ is the Fourier transform of $\lambda_h(z)$. Consequently, we find
\begin{equation}
    m_1=\frac{c\sigma_{z0}}{\pi} \int_0^\infty \text{Re}[Z_\parallel(k)] e^{-k^2\sigma_{z0}^2} dk
    \label{eq:m1}
\end{equation}
with the impedance property $Z_\parallel^*(k)=Z_\parallel(-k)$~\cite{chao1993physics}.

\subsubsection{Peak position of bunch profile}
The peak position of the bunch profile (that is, the relative maximum of $\lambda_h(z)$) is of interest from a measurement point of view. For example, the beam position monitors (BPMs) in storage rings monitor the beam's electromagnetic fields and output a signal with its voltage roughly proportional to $d\lambda_h(z)/dz$ (for example, see~Ref.~\cite{ieiri2009beam}). Detecting the zero-crossing point of a BPM signal can give information on the peak position of the bunch profile. Another way is to extract the peak position from the bunch profile measured by a streak camera (SC, see Sec.~7.4.8 of~\cite{chao2023handbook} for further details and~\cite{carver2023beam} for a successful application of SC).

From Eq.~(\ref{eq:de_Haissinski}), taking $d\lambda_h(z)/dz=0$ yields the peak position of the bunch profile
\begin{equation}
    z_m(I)=I\sigma_{z0}\mathcal{W}_\parallel(z_m).
    \label{eq:peak}
\end{equation}
The bunch profile can have single or multiple peaks depending on the impedance properties and the bunch current. For simplicity of discussion, here we assume only one peak for a given bunch profile. Given machine parameters and an impedance model, the current-dependent bunch profile and the corresponding wake potential can be obtained by simulations. Equation~(\ref{eq:peak}) shows how the bunch profile's peak and the corresponding wake potential should be correlated. Therefore, comparing the simulated and measured $z_m$ as a function of $I$ provides another way to verify the impedance model for a storage ring.  

The quantities $z_c$ and $z_m$ are identical when the profile of the bunch is symmetric with respect to $z_c$, which is the case for a Gaussian bunch in electron storage rings. The Haissinski equation with a purely inductive impedance also has identical $z_c$ and $z_m$~\cite{thomas2002analytical}. However, in cases where the real part of the impedance is nonzero, the bunch profile of~(\ref{eq:Haissinski_equation_explicit}) is asymmetric, leading to different scaling laws for $z_c(I)$ and $z_m(I)$, as illustrated in the following analysis.

The peak shift rate at $I=0$ is given by
\begin{equation}
    m_2\equiv \left . \frac{dz_m}{dI} \right |_{I=0}=\sigma_{z0}\mathcal{W}_\parallel(0).
\end{equation}
In terms of impedance, there is
\begin{equation}
    m_2=\frac{c\sigma_{z0}}{\pi} \int_0^\infty \text{Re}[Z_\parallel(k)] e^{-k^2\sigma_{z0}^2/2} dk.
    \label{eq:m2}
\end{equation}
Equations~(\ref{eq:m1}) and~(\ref{eq:m2}) show that the shift rates $m_1$ and $m_2$ depend only on the real part of the impedance for a Gaussian bunch. Measurement of them will be beneficial in understanding the resistive impedance of a storage ring. Since $\text{Re}[Z_\parallel(k)]\geq 0$ for any $k$, there is $m_2>m_1$, suggesting that the profile peak is shifting faster than the center of mass when the bunch current increases. In particular, considering a purely resistive impedance $Z_\parallel(k)=R$, there is $m_2=\sqrt{2}m_1$. 

\subsubsection{Potential-well bunch lengthening}
The preceding subsections have demonstrated the sensitivity of the centroid shift and the peak position of the circulating bunch to the real part of the impedance. The lengthening of the bunch in a storage ring is mainly attributed to the influence of the imaginary impedance~\cite{zotter1981potential}. Given the bunch profile $\lambda_h(z)$, the rms bunch length can be calculated by
\begin{equation}
    \sigma_z^2=\int_{-\infty}^\infty
    (z-z_c)^2\lambda_h(z)dz
    =\int_{-\infty}^\infty
    z^2\lambda_h(z)dz-z_c^2,
\end{equation}
with $z_c$ the center of mass previously defined. We show how to calculate $\sigma_z$ from Eq.~(\ref{eq:de_Haissinski}). Multiplying $z$ on both sides of Eq.~(\ref{eq:de_Haissinski}) and performing integration over $z$, we can obtain three terms. The first term is a constant -1 (considering that $\lambda_h(z)$ decays exponentially as $e^{-z^2/(2\sigma_{z0}^2)}$ when $z\rightarrow\pm \infty$, according to the Haissinski equation). The second term equals $(\sigma_z^2+z_c^2)/\sigma_{z0}^2$. The third term is an integration that contains the wake function. Combining the three terms, we can arrive at an equation that describes the potential-well bunch lengthening
\begin{equation}
    x^2-1-\frac{cI}{2\pi\sigma_{z0}} Z_\parallel^\text{eq}(x)=0,
    \label{eq:PW_Lengthening_from_Haissinski}
\end{equation}
where $x=\sigma_z/\sigma_{z0}$ is the bunch lengthening factor and the term $Z_\parallel^\text{eq}(x)$ is formulated by
\begin{equation}
    Z_\parallel^\text{eq}(x) =
    \frac{2\pi}{c} \int_{-\infty}^\infty dz (z-z_c) \lambda_h(z)
    \mathcal{W}_\parallel(z).
    \label{eq:Zeff1}
\end{equation}
In terms of impedance, Eq.~(\ref{eq:Zeff1}) is equivalent to
\begin{equation}
        Z_\parallel^\text{eq}(x)=
        -\int_{-\infty}^\infty dk Z_\parallel(k) \tilde{\lambda}_h(k)
        \left[ i\frac{d}{dk}\tilde{\lambda}_h^*(k) +
        z_c \tilde{\lambda}_h^*(k) \right].
    \label{eq:Zeff2}
\end{equation}
Here, we define $Z_\parallel^\text{eq}(x)$ as an equivalent impedance expressed as a function of bunch lengthening factor, although its determination is indeed influenced by the bunch profile. Note that this equivalent impedance is always real but not complex. As a straightforward corollary of the Haissinski equation,  Eq.~(\ref{eq:PW_Lengthening_from_Haissinski}) shows that the term $Z_\parallel^\text{eq}$ is simply a quadratic function of $x$, while $x$ obviously is a function of the normalized current $I$ (that is, $x=x(I)$). Equation~(\ref{eq:Zeff2}) indicates that when the density distribution is deformed, the real part of the impedance contributes to the bunch lengthening, although the imaginary part is usually the dominant source. The reader may notice a similarity of the equivalent impedance $Z_\parallel^\text{eq}$ defined here with the conventional effective impedance $(Z_0^\parallel/\omega)_\text{eff}$, which measures the shift in the complex mode frequencies, for the instability theory in the storage rings~\cite{sacherer1977bunch, chao1993physics}. In fact, by expanding the density spectrum $\tilde{\lambda}_h(k)$ into the sum of azimuthal and radial modes~\cite{sacherer1977bunch, cai2011linear}, our formulation can be connected to the conventional formulation of effective impedance. However, the details will be reserved for a later paper.

At $I=0$, the density distribution is Gaussian with $z_c=0$ and $x(0)=1$. The bunch lengthening rate at $I=0$ is given by
\begin{equation}
    m_3\equiv \left . \frac{dx}{dI} \right |_{I=0}=\frac{c}{4\pi\sigma_{z0}}Z_{\parallel 0}^\text{eq}.
\end{equation}
Here, the equivalent impedance at zero current $Z_{\parallel 0}^\text{eq}\equiv Z_\parallel^\text{eq}(1)$ is given by Eq.~(\ref{eq:Zeff2}) with $z_c=0$ and $\tilde{\lambda}_h(k)=e^{-k^2\sigma_{z0}^2/2}$. It depends only on the imaginary part of $Z_\parallel(k)$, that is,
\begin{equation}
    Z_{\parallel 0}^\text{eq}=-2\sigma_{z0}^2
    \int_0^\infty dk k \text{Im}[Z_\parallel(k)] e^{-k^2\sigma_{z0}^2},
    \label{eq:Zeff3}
\end{equation}
suggesting that the inductive part of the impedance solely determines the lengthening rate of bunch length at zero current. Though Eq.~(\ref{eq:PW_Lengthening_from_Haissinski}) shows a simple relation of $Z_\parallel^\text{eq}$ with the bunch lengthening factor $x$, the relation between $Z_\parallel^\text{eq}$ and the normalized current $I$ is complicated. Here, we only give the slope at $I=0$ as $dZ_\parallel^\text{eq}/dI|_{I= 0}=\frac{2\pi\sigma_{z0}}{c}\left[ m_3^2 + x'' \right]$ with $x''=d^2x/dI^2$ to be determined.

The quantity $Z_{\parallel 0}^\text{eq}$ can be taken as an effective inductance (see Sec.~\ref{sec:ExplicitExpressions} for further discussion), which can be computed with the nominal bunch length using the numerically constructed impedance model. When validating an impedance budget that gives $Z_{\parallel 0}^\text{eq}$(theory) for a storage ring, the process involves measuring $Z_{\parallel}^\text{eq}$ (that is, measuring the bunch length) as a function of $I$ and then extrapolating its slope at $I=0$ to determine $m_3$, thus obtaining $Z_{\parallel 0}^\text{eq}$(measurement). The difference between $Z_{\parallel 0}^\text{eq}$(theory) and $Z_{\parallel 0}^\text{eq}$(measurement) can quantify the gap between the theoretical model and the actual impedance of a machine in the imaginary part.

\subsubsection{Inverse problem of Haissinski equation}
Given a bunch profile obtained from simulations or beam-based measurements, one may be interested in extracting the impedance from Eq.~(\ref{eq:Haissinski_equation_explicit}). This leads to the inverse problem of the Haissinski equation. From Eq.~(\ref{eq:Haissinski_equation_explicit}), the wake potential can be calculated from the line density as
\begin{equation}
    \mathcal{W}_\parallel(z)=\frac{\sigma_{z0}}{I}
    \left[ \frac{d\ln\lambda_h(z)}{dz} + \frac{z}{\sigma_{z0}^2} \right].
    \label{eq:WakePotentialhai}
\end{equation}
Taking Fourier transform to both sides of the above equation, one can calculate the impedance from the wake potential as
\begin{equation}
    Z_\parallel(k)=
    \frac{1}{c\tilde{\lambda}_h(k)}
    \int_{-\infty}^\infty \mathcal{W}_\parallel(z) e^{-ikz} dz.
    \label{eq:InvHaissinski3}
\end{equation}
In terms of the bunch profile, there is~\cite{chao2020lectures}
\begin{equation}
    Z_\parallel(k)=
    \frac{\sigma_{z0}}{Ic\tilde{\lambda}_h(k)}
    \int_{-\infty}^\infty \left[ \frac{d\ln\lambda_{0}(z)}{dz}+\frac{z}{\sigma_{z0}^2} \right] e^{-ikz} dz.
    \label{eq:InvHaissinski1}
\end{equation}
Performing integration by parts, the above equation can be rewritten as
\begin{equation}
    Z_\parallel(k)=
    \frac{ik\sigma_{z0}}{Ic\tilde{\lambda}_h(k)}
    \int_{-\infty}^\infty \left[ \ln\lambda_{0}(z)+\frac{z^2}{2\sigma_{z0}^2} \right] e^{-ikz} dz.
    \label{eq:InvHaissinski2}
\end{equation}
In terms of potential well, it can be rewritten as
\begin{equation}
    Z_\parallel(k)=
    \frac{ik\sigma_{z0}}{Ic\tilde{\lambda}_h(k)}
    \int_{-\infty}^\infty \left[ -V(z)+\frac{z^2}{2\sigma_{z0}^2} \right] e^{-ikz} dz.
    \label{eq:InvHaissinski4}
\end{equation}
The above equations show, mathematically, it is feasible to calculate the frequency-dependent impedance if the beam distribution is accurately obtained~\cite{chao2020lectures}.

Calculating the wake potential from the deformed density distribution using Eq.~(\ref{eq:WakePotentialhai}) is relatively straightforward. However, extraction of the impedance from the wake potential is known to be a deconvolution problem. Due to the inherent characteristics of deconvolution problems, the inverse problem associated with the Haissinski equation is mathematically well-defined but ill-posed. The problem is sensitive to errors that are always present in the simulated or measured data for $\lambda_h(z)$. Practically, it can be challenging to retrieve accurate impedance data at frequencies $k\gg 1/\sigma_{z0}$ using Eq.~(\ref{eq:InvHaissinski1}) or (\ref{eq:InvHaissinski2}). However, if the bunch profile is reliable, it is possible to obtain impedance data with good accuracy at frequencies $k\lesssim 1/\sigma_{z0}$.


\section{\label{sec:ExplicitExpressions}Explicit expressions for specific impedances}

For some well-defined impedances, such as pure inductance $L$, pure resistance $R$, and pure capacitance $C$, the equivalent impedance can be explicitly formulated as follows.

For a purely inductive impedance $Z_\parallel(k)=-ikcL$, there are $W_\parallel(z)=-c^2L\delta'(z)$, $z_c=0$~\cite{thomas2002analytical}, and
\begin{equation}
    Z_\parallel^\text{eq}=\pi cL \int_{-\infty}^\infty \lambda_h^2(z) dz
\end{equation}
with the Dirac delta function $\delta(z)$.

For a purely resistive impedance $Z_\parallel(k)=R$, there are $W_\parallel(z)=cR\delta(z)$ and
\begin{equation}
    Z_\parallel^\text{eq}=2\pi R \int_{-\infty}^\infty (z-z_c)\lambda_h^2(z) dz.
\end{equation}

For a purely capacitive impedance of
\begin{equation}
    Z_\parallel(k)=\frac{i}{kcC},
    \label{eq:capacitiveZ}
\end{equation}
there is
\begin{equation}
    Z_\parallel^\text{eq}=\frac{2\pi}{cC} \int_{-\infty}^\infty (z-z_c)\lambda_h(z) \Lambda_0(z) dz
\end{equation}
with $\Lambda_0(z)=\int_z^\infty \lambda_h(z)dz$ an accumulation function. The concept of capacitive impedance was first introduced in~\cite{bane1990bunch}. Note that Eq.~(\ref{eq:capacitiveZ}) is not equivalent to the conventionally defined purely capacitive impedance of a causal wake function. For further discussion, see Appendix~\ref{sec:capacitor}.

The Haissinski distributions for the aforementioned impedances have been well investigated in the literature~\cite{ruggiero1977theory, zotter1990review, shobuda1998existence, shobuda2001proof}. They can be used as input to calculate the corresponding equivalent impedance $Z_\parallel^\text{eq}$ numerically. For an absolute impedance, one has to solve the Haissinski equation numerically (see~\cite{warnock2018numerical} and references therein), obtain the equilibrium distribution $\lambda_h(z)$, and then calculate the equivalent impedance.

\begin{table*}[ht]
\caption{Equivalent impedances and centroid shifts for specific impedance forms with a Gaussian approximation of Haissinski distribution.}
\centering
\begin{tabular}{|p{3cm}|p{4.4cm}|p{5.5cm}|p{3.8cm}|}
\hline
\multirow{2}{*}{Description} & \multirow{2}{*}{Impedances $Z_\parallel(k)$} & \multirow{2}{*}{Equivalent impedance $Z_\parallel^\text{eq}(x)$} & \multirow{2}{*}{Centroid shift $z_c(x)$} \\
& & & \\
\hline
\multirow{2}{*}{Pure inductance} & \multirow{2}{*}{$-ikcL$} & \multirow{2}{*}{$\frac{\sqrt{\pi}cL}{2\sigma_{z0}x}$} & \multirow{2}{*}{0} \\
& & & \\
\hline
\multirow{2}{*}{Pure resistance} & \multirow{2}{*}{$R$} & \multirow{2}{*}{0} & \multirow{2}{*}{$\frac{IcR}{2\sqrt{\pi}x}$} \\
& & & \\
\hline
\multirow{2}{*}{Pure capacitance} & \multirow{2}{*}{$\frac{i}{kcC}$} & \multirow{2}{*}{$-\frac{\sqrt{\pi}\sigma_{z0}x}{cC}$} & \multirow{2}{*}{0} \\
& & & \\
\hline
\multirow{2}{*}{Resistive wall} & \multirow{2}{*}{$\frac{L}{2\pi b}
    \left[ 1-i \text{sgn}[k]\right]
    \sqrt{\frac{|k|Z_0}{2\sigma_c}}$} & \multirow{2}{*}{$\frac{L}{2\pi b}\sqrt{\frac{Z_0}{2\sigma_c}} \frac{\Gamma\left( \frac{5}{4} \right)}{\sqrt{\sigma_{z0}x}}$} & \multirow{2}{*}{$\frac{L}{2\pi b}\sqrt{\frac{Z_0}{2\sigma_c}} \frac{cI\Gamma\left( \frac{3}{4} \right)}{\pi\sqrt{\sigma_{z0}}x^{3/2}}$} \\
& & & \\
\cline{2-4}
& \multicolumn{3}{l|}{$L$: chamber length; $b$: chamber radius; $\sigma_c$: Conductivity.} \\
\hline
\multirow{2}{*}{Steady-state CSR} \newline \multirow{2}{*}{in free-space} & \multirow{3}{*}{$\frac{Z_0}{3^{1/3}}\left( \frac{\sqrt{3}}{2}+\frac{1}{2}i \right)\Gamma \left( \frac{2}{3} \right) (k\rho)^{1/3}$} & \multirow{3}{*}{$-\frac{Z_0\Gamma\left(\frac{2}{3}\right)\rho^{1/3}}{2\cdot 3^{1/3}} \frac{\Gamma\left(\frac{7}{6}\right)}{(\sigma_{z0}x)^{1/3}}$} & \multirow{3}{*}{$\frac{3^{1/6}\Gamma^2\left(\frac{2}{3}\right)}{4\pi} \frac{cIZ_0\rho^{1/3}}{\sigma_{z0}^{1/3}x^{4/3}}$} \\
& & & \\
\cline{2-4}
& \multicolumn{3}{l|}{$\rho$: bending radius; assume $2\pi\rho$ for the total length of dipoles.} \\
\hline
\multirow{2}{*}{Steady-state CWR} \newline \multirow{2}{*}{in free-space~\cite{stupakov2016analytical}} & \multirow{3}{*}{$\frac{1}{16}Z_0\theta_0^2Lk\left(1-\frac{2i}{\pi}
    \ln \frac{k}{k_c} \right)$} & \multirow{3}{*}{$-\frac{Z_0\theta_0^2L}{32\sqrt{\pi}x\sigma_{z0}}
    \left[ Y + 2 \ln \left( k_cx\sigma_{z0} \right) \right]$} & \multirow{3}{*}{$\frac{cIZ_0\theta_0^2L}{32\pi\sigma_{z0}x^2}$} \\
& & & \\
\cline{2-4}
& \multicolumn{3}{l|}{$L$: wiggler length; $\theta_0$: wiggler deflection angle; $k_c$: fundamental frequency} \\
& \multicolumn{3}{l|}{of wiggler radiation; $Y=-2 + \gamma_E +\ln 4\approx -0.0365$.} \\
\hline
\multirow{2}{*}{Resonator model} \newline \multirow{2}{*}{($Q>1/2$)} & \multirow{3}{*}{$\frac{R_s}{1+iQ(k_r/k-k/k_r)}$} & \multirow{3}{*}{$-\frac{\pi R_s \sigma_z}{Q'}\left[ \frac{k_rQ'}{\sqrt{\pi}Q} + \sigma_z\text{Im}[k_1^2w(k_1\sigma_z)] \right]$} & \multirow{3}{*}{$\frac{cI\sigma_{z0}R_s}{2Q'}\text{Re}[k_1w(k_1\sigma_z)]$} \\
& & & \\
\cline{2-4}
& \multicolumn{3}{l|}{$R_s$: shunt impedance; $Q$: quality factor; $k_r$: resonant frequency; $Q'=\sqrt{Q^2-1/4}$;} \\
& \multicolumn{3}{l|}{$k_1=\frac{k_r}{Q}[-i/2+Q']$; $w(z)=e^{-z^2}\left[1-i\text{Erfi}(z)\right]$; $\text{Erfi}(z)$: imaginary error function; $\sigma_z=\sigma_{z0}x$.} \\
\hline
\end{tabular}
\label{tb:zeff_summary}
\end{table*}
Further calculations can be performed when a Gaussian distribution with rms length $\sigma_z$ and center of mass $z_c$ is used to approximate the Haissinski distribution. Then, the equivalent impedance is explicitly written as follows, from Eq.~(\ref{eq:Zeff2}): 
\begin{equation}
    Z_\parallel^\text{eq}=-2\sigma_z^2
    \int_0^\infty dk k \text{Im}[Z_\parallel(k)] e^{-k^2\sigma_z^2}.
    \label{eq:Zeff4}
\end{equation}
In this case, only the imaginary part of the raw impedance contributes to the equivalent impedance. For some well-established impedance models in the literature, the explicit formulas for equivalent impedance and centroid shift are summarized in Table~\ref{tb:zeff_summary}. They can be applied to Eq.~(\ref{eq:PW_Lengthening_from_Haissinski}) to predict the potential-well bunch lengthening. In particular, for a purely inductive impedance, it gives the popular Zotter's cubic equation:
\begin{equation}
    x^3-x-D=0
    \label{eq:cubic_eq}
\end{equation}
with
\begin{equation}
    D=\frac{c^2IL}{4\sqrt{\pi}\sigma_{z0}^2}
    =\frac{cI_bL}{4\sqrt{\pi}\eta\sigma_{z0}\sigma_\delta^2(E/e)},
\end{equation}
with $I_b=Nec/C_0$ being the single bunch current. Here, we show that the numerical constant in the denominator of $D$ is exactly $4\sqrt{\pi}$, as suggested by some authors (see~\cite{smaluk2018impedance, zhou2023potential} and references therein), but not $\sqrt{2\pi}$ as formulated in~\cite{zotter1981potential}. We also demonstrate that the Zotter's cubic equation is only a special case of our generalized quadratic equation~(\ref{eq:PW_Lengthening_from_Haissinski}) with the assumptions 1) a Gaussian distribution well approximates the Haissinski distribution and 2) a purely inductive impedance well describes the total impedance of the storage ring.

When inductive impedance dominates total impedance, the cubic scaling law Eq.~(\ref{eq:cubic_eq}) fairly fits measurements of bunch lengthening. This is true for many modern electron storage rings and justifies using the cubic equation to estimate effective inductance $L_\text{eff}$~\cite{bane1990bunch} from the bunch length measurements~\cite{smaluk2018impedance}. By connecting Eqs.~(\ref{eq:Zeff4}) and (\ref{eq:cubic_eq}), for an absolute impedance, the effective inductance at zero bunch current can be calculated by
\begin{equation}
    L_\text{eff}=\frac{2\sigma_{z0}}{\sqrt{\pi}c}Z_{\parallel 0}^\text{eq}.
    \label{eq:Leff}
\end{equation}
This formulation is equivalent to Eq.~(5) of Ref.~\cite{bane1990bunch}, and Eq.~(5) of Ref.~\cite{smaluk2018impedance} where $h(\omega)$ should be replaced by $h_1(\omega)=(\omega\sigma_{z0}/c)^{2}e^{-\omega^2\sigma_{z0}^2/c^2}$ as suggested in~\cite{zotter1981potential}.

We emphasize that Eq.~(\ref{eq:cubic_eq}) implicitly assumes that the lengthening of the bunch in a storage ring is fully attributed to pure inductance. In general, this is not the case and may pose a challenge when comparing the experimental results with numerical computations of impedance effects. An improvement to the cubic equation Eq.~(\ref{eq:cubic_eq}) is to rewrite $D$ as $D(I)=IG(I)$. Measurement of $G(I)=(x^3-x)/I$ through measurement of bunch lengths determines the impedance characteristics of a storage ring. Especially when the inductive impedance dominates the lengthening of the bunch, $G(I)$ should show a weak dependence on the normalized current $I$.

\section{\label{sec:applications}Applications to real machines}

For a modern and complicated storage ring, a ``bottom-up'' approach to a pseudo-Green function wake (for examples of this approach, see~\cite{bane2010pep, warnock2018numerical}) is usually used to obtain the broad-band impedance model for studies of impedance effects. A Gaussian bunch with length $\hat{\sigma}_z$ significantly shorter than the nominal bunch length $\sigma_{z0}$ is used as the driving bunch in numerical calculations. Wake fields are monitored at long distances up to $s_\text{max}$. Typically, the ratio $\sigma_{z0}/\hat{\sigma}_z$ is chosen as 5-10, depending on the available computing resources. And $s_\text{max}/\sigma_{z0}$ is chosen to be larger than 10 for simulations of single-bunch impedance effects. The pseudo-Green function wakes from individual components are summed up to construct a total wake for later simulations of beam instabilities. In this section, we consider a few examples of storage rings to test the theories formulated in the previous sections and discuss the reflections to beam-based impedance measurements. The reader may refer to~\cite{zhou2023potential} for another relevant example of SuperKEKB high energy ring.

\subsection{SLC damping ring}

\begin{figure}
\centering
\includegraphics[width=0.85\linewidth]{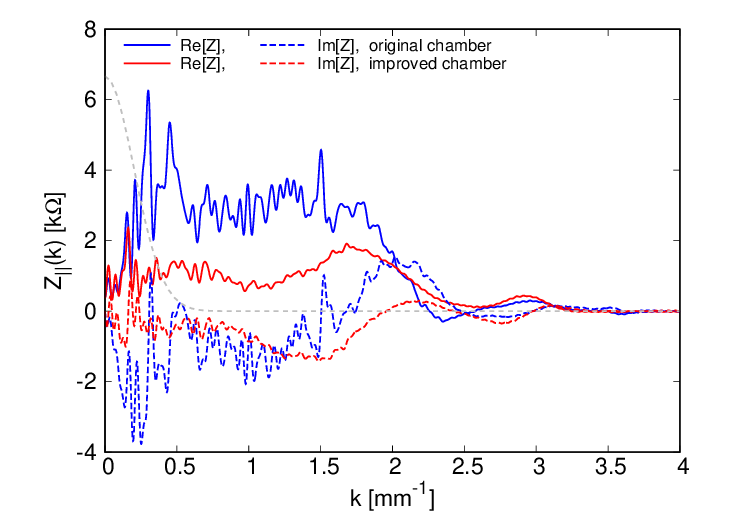}
\caption{Impedances calculated from the pseudo-green wake functions for SLC damping rings. The blue and red lines indicate results for the original and improved vacuum chambers, respectively. The gray dashed line shows the beam spectrum of the nominal Gaussian bunch with $\sigma_{z0}=4.95$ mm.}
\label{fig:impedanceSLC-DR}
\end{figure}

\begin{figure}
\centering
\includegraphics[width=0.85\linewidth]{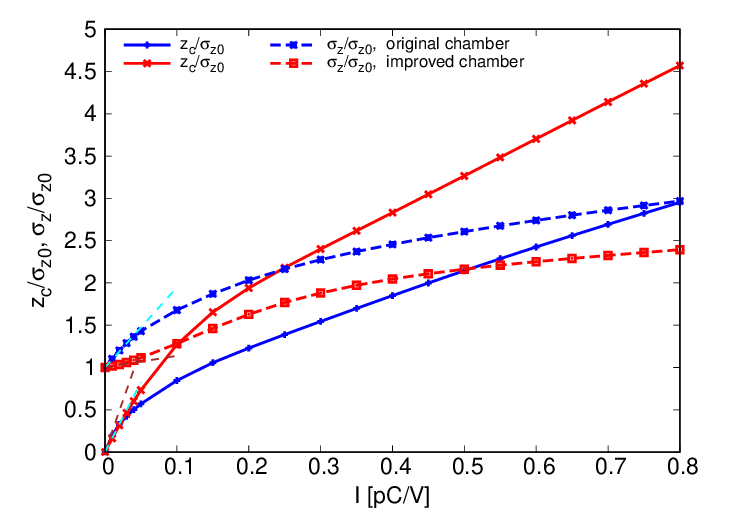}
\caption{Bunch centroid and rms bunch length as a function of normalized current for SLC damping rings. The dashed straight lines are the slopes at zero current given by $m_1$ and $m_3$.}
\label{fig:sigzSLC-DR}
\end{figure}

\begin{figure}
\centering
\includegraphics[width=0.85\linewidth]{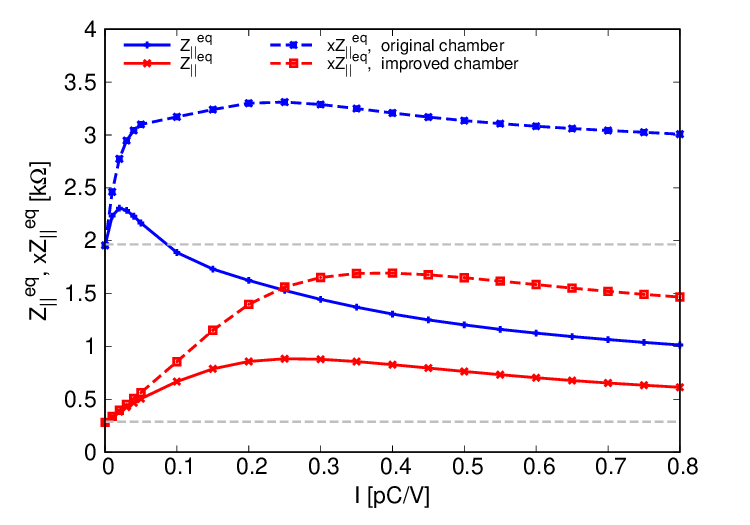}
\caption{Equivalent impedance as a function of normalized current calculated from simulated bunch lengths for SLC damping rings. The gray dashed lines show the impedance data corresponding to Zottter's cubic equation.}
\label{fig:ZeffSLC-DR}
\end{figure}

We continue the work of Warnock and Bane in~\cite{warnock2018numerical}, where the numerical calculations of the Haissinski solution for the damping rings (DR) of the Stanford Linear Collider (SLC) were carefully studied, to test the formulations in Sec.~\ref{sec:theories}. For SLC DR with the original and improved vacuum chamber, the beam parameters are summarized in Table~\ref{tb:SLC-DR-parameters}. For these two configurations, the pseudo-Green wake functions are shown in Figs. 1 and 4 of~\cite{warnock2018numerical}. The corresponding impedances are calculated using the chirp Z-transform~\cite{rabiner1969chirp} (CZT, a generalization of the discrete Fourier transform (DFT)). The results are shown in Fig.~\ref{fig:impedanceSLC-DR}. The CZT is used here to achieve a better frequency resolution than the DFT, especially for low frequencies. The centroid shift and bunch lengthening are obtained by numerically solving the Haissinski equation. The results are shown in Fig.~\ref{fig:sigzSLC-DR}, consistent with Figs. 3 and 6 of~\cite{warnock2018numerical}. Then, from bunch length, the equivalent impedances $Z_\parallel^\text{eq}$ as a function of the normalized bunch currents are calculated using Eq.~(\ref{eq:PW_Lengthening_from_Haissinski}), as shown in Fig.~\ref{fig:ZeffSLC-DR}.
\begin{table}[hbt]
   \centering
   \caption{Beam parameters for SLC damping rings. Equivalent impedance, effective inductance and loss factor are respectively calculated using Eqs.(\ref{eq:Zeff4}), (\ref{eq:Leff}), and~(\ref{eq:LossFactor}) with $I=0$.}
   \begin{tabular}{lcc}
       \toprule
       \textbf{Parameter} & \textbf{Original}                      & \textbf{Improved} \\
           Beam energy (GeV)       & 1.15            & 1.15        \\
           Nominal bunch length (mm)       & 4.95            & 4.95       \\
           Synchrotron tune         & -0.0117      & -0.0116 \\
           Long. damping time (ms)     & 1.7   & 1.7 \\
           Energy spread ($10^{-4}$)   & 7.0     & 7.4 \\
           Equivalent impedance ($\Omega$)   & 1970     & 290 \\
           Effective inductance (nH)   & 36.6     & 5.3 \\
           Loss factor (V/pC) & 24.4     & 16.5 \\
           \hline
   \end{tabular}
   \label{tb:SLC-DR-parameters}
\end{table}

Using impedance data, the equivalent impedance, the effective inductance, and the loss factor at zero current are also calculated, as shown in Table~\ref{tb:SLC-DR-parameters}. From these parameters and Fig.~\ref{fig:impedanceSLC-DR}, one can see that the smoothing of the chamber reduced the total inductance by 85\%. Meanwhile, the total loss factor was reduced by about 30\%. This change converted the SLC DR from an inductance-dominant ring to a resistance-dominant ring~\cite{warnock2018numerical}. Thus, the bunch lengthening became weaker, but the centroid shift became stronger, as seen in Fig.~\ref{fig:sigzSLC-DR}. The relatively large resistive impedance of SLC DR also plays an important role in the bunch lengthening, as shown in Fig.~\ref{fig:ZeffSLC-DR}. In this figure, we plot the impedance data of $Z_\parallel^\text{eq}$ and $xZ_\parallel^\text{eq}$, which correspond to our generalized quadratic equation and Zotter's cubic equation, respectively. If the inductive impedance dominates the total impedance, then $xZ_\parallel^\text{eq}$ should be constant or have a weak dependence on the bunch current, as shown in the gray dashed line of Fig.~\ref{fig:ZeffSLC-DR}. However, $xZ_\parallel^\text{eq}$ is a nonlinear function of $I$, indicating a complicated interplay between the real and imaginary parts of the impedance in the SLC DR. This interplay appears in the extra terms containing real-part impedances in Eq.~(\ref{eq:Zeff2}) due to potential-well distortion, which creates imaginary part in the beam spectrum.

\subsection{KEKB low energy ring}

\begin{figure}
\centering
\includegraphics[width=0.85\linewidth]{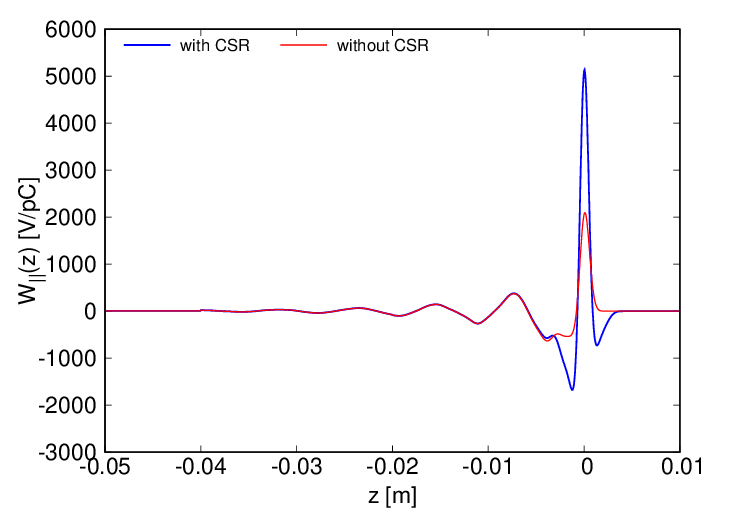}
\includegraphics[width=0.85\linewidth]{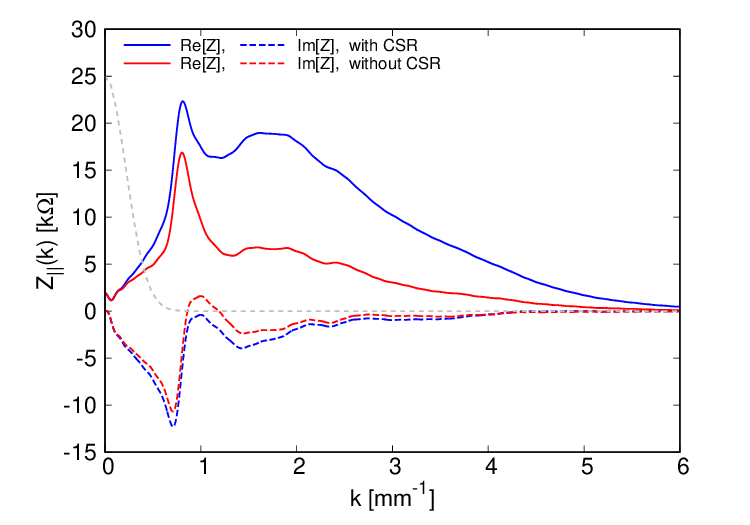}
\caption{Pseudo-Green function wakes and corresponding impedances for KEKB LER. The gray dashed line shows the beam spectrum of the nominal Gaussian bunch with $\sigma_{z0}=4.58$ mm.}
\label{fig:impedanceKEKB-LER}
\end{figure}

\begin{figure}
\centering
\includegraphics[width=0.85\linewidth]{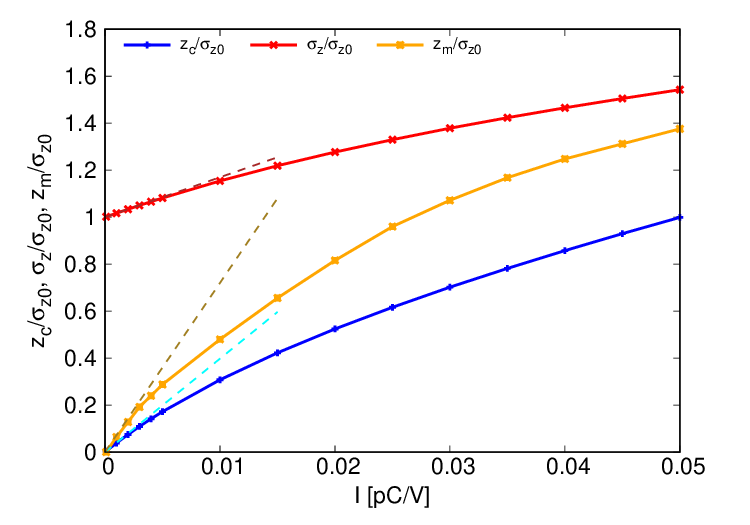}
\caption{Bunch centroid, rms bunch length and peak position as a function of normalized current for KEKB LER. The data are obtained from the numerically solved Haissinski equation using the impedance model including CSR of Fig.~\ref{fig:impedanceKEKB-LER}. The dashed lines are the slopes at zero current given by $m_1$, $m_2$, and $m_3$.}
\label{fig:sigzKEKB-LER}
\end{figure}

\begin{figure}
\centering
\includegraphics[width=0.85\linewidth]{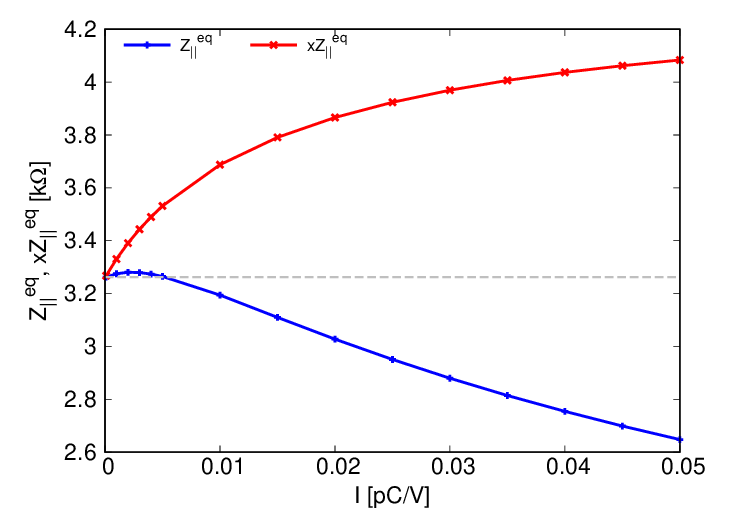}
\caption{Equivalent impedance as a function of normalized current calculated from simulated bunch lengths for KEKB LER. The gray dashed line shows the equivalent impedance corresponding to Zottter's cubic equation.}
\label{fig:ZeffKEKB-LER}
\end{figure}

The KEKB low energy ring (LER) case was reviewed in~\cite{zhou2009simulation, warnock2018numerical}. The main beam parameters are summarized in Table~\ref{tb:LER-parameters}. The longitudinal wakes for various components have been calculated using a Gaussian driving bunch with length $\hat{\sigma}_z$=0.5 mm and summed to create the pseudo-Green function wake~\cite{zhou2009simulation} as shown in Fig.~\ref{fig:impedanceKEKB-LER}. The longitudinal impedances of coherent synchrotron radiation (CSR) from dipoles and damping wigglers are included in the impedance model because their effects are non-negligible due to the relatively small bending radius of about 16 m and the large chamber radius of 47 mm. Bunch centroid, peak position, and rms length as a function of normalized current are shown in Fig.~\ref{fig:sigzKEKB-LER}. The corresponding equivalent impedance is shown in Fig.~\ref{fig:ZeffKEKB-LER}. It is seen that the bunch lengthening cannot be well described by the cubic equation.

\begin{table}[hbt]
   \centering
   \caption{Beam parameters for LER of KEKB and SuperKEKB. Equivalent impedance, effective inductance, and loss factor are respectively calculated using Eqs.(\ref{eq:Zeff4}), (\ref{eq:Leff}), and~(\ref{eq:LossFactor}) with $I=0$.}
   \begin{tabular}{lcc}
       \toprule
       \textbf{Parameter} & \textbf{KEKB}                      & \textbf{SuperKEKB} \\
           Beam energy (GeV)       & 3.5            & 4.0        \\
           Nominal bunch length (mm)       & 4.58            & 4.60       \\
           Synchrotron tune         & -0.024      & -0.0233 \\
           Long. damping time (ms)     & 21.6   & 22.8 \\
           Energy spread ($10^{-4}$)   & 7.27     & 7.53 \\
           Equivalent impedance ($\Omega$)   & 3260     & 1990 \\
           Effective inductance (nH)   & 56.8     & 34.5 \\
           Loss factor (V/pC) & 40.0     & 24.7 \\
           \hline
   \end{tabular}
   \label{tb:LER-parameters}
\end{table}

An issue in KEKB LER was that the measured bunch lengthening using a streak camera was much more severe than the predictions of simulations using the constructed impedance model. To reproduce the measurements, a pure inductance of around 90 nH was required~\cite{zhou2009simulation}. However, this value is very high compared to the effective inductance calculated from the impedance model shown in Table~\ref{tb:LER-parameters}. To resolve this discrepancy, two different methods were used to compare the predictions of the impedance model. One was to measure the beam phase using a gated beam position monitor (BPM)~\cite{ieiri2009beam}, that is, to measure the phase difference between the signal from the BPM to the reference RF signal. Another was to extrapolate the beam power due to wakefields from the klystron power measured by the RF system. The detailed results are presented in~\cite{zhou2017icfa}. It was found that the beam-phase measurements using gated BPM agree well with the theoretical calculations. From the klystron power measurements, the beam phase was extracted and good agreement was also found with calculations based on the impedance model. These facts suggest that the ring broad-band impedance was reliable, while systematic errors could not be ruled out from the streak camera measurements.

\subsection{SuperKEKB low energy ring}

\begin{figure}
\centering
\includegraphics[width=0.85\linewidth]{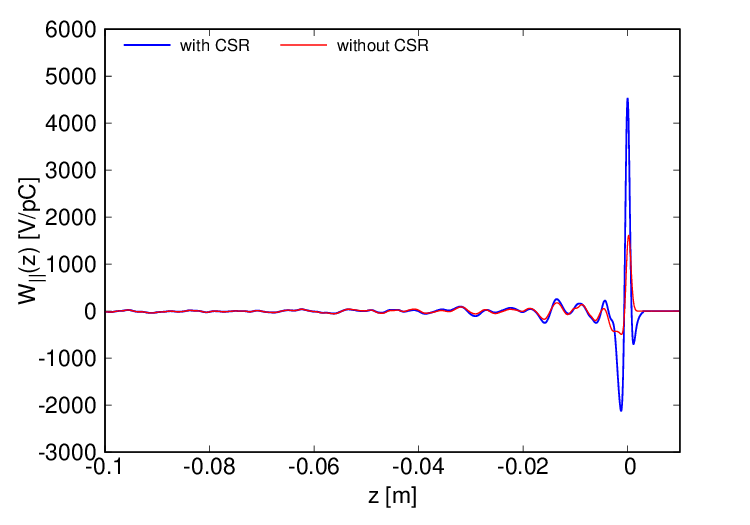}
\includegraphics[width=0.85\linewidth]{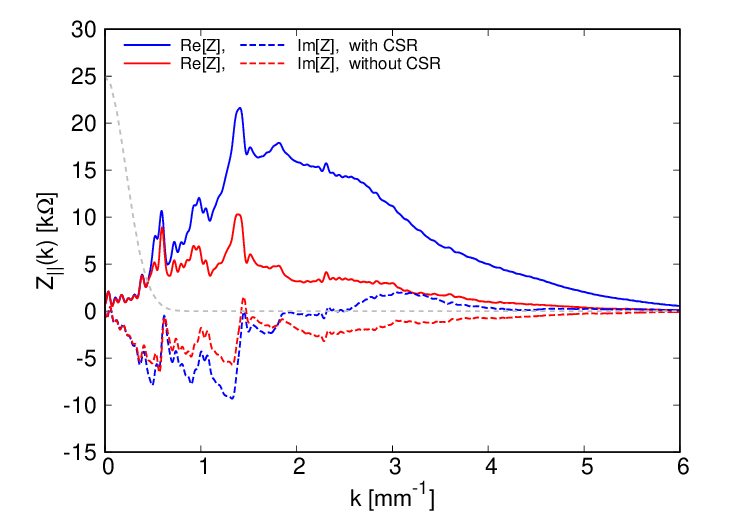}
\caption{Pseudo-Green function wakes and corresponding impedances for SuperKEKB LER. The gray dashed line shows the beam spectrum of the nominal Gaussian bunch with $\sigma_{z0}=4.6$ mm.}
\label{fig:wakeSKBLER}
\end{figure}

\begin{figure}
\centering
\includegraphics[width=0.85\linewidth]{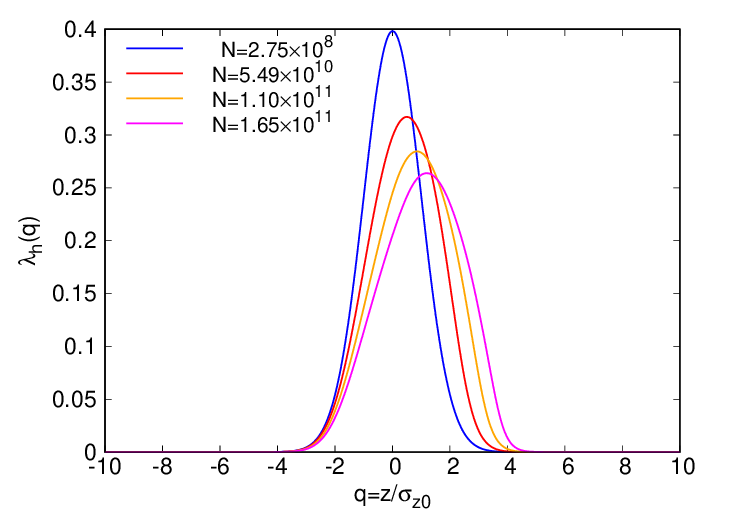}
\caption{Haissinski solution for SuperKEKB LER with $N=(0.0275, 5.49, 11.0, 16.5)\times 10^{10}$.}
\label{fig:haiDensity}
\end{figure}

\begin{figure}
\centering
\includegraphics[width=0.85\linewidth]{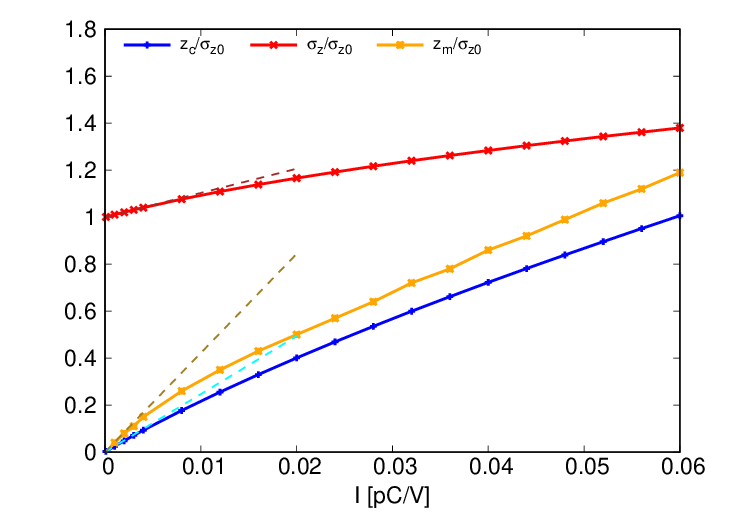}
\caption{Bunch centroid, peak position, and rms bunch length as a function of normalized current for SuperKEKB LER. The dashed lines are the slopes at zero current given by $m_1$, $m_2$, and $m_3$.}
\label{fig:sigz}
\end{figure}

For SuperKEKB LER, the longitudinal wakes of various components have been carefully calculated using a Gaussian driving bunch with length $\hat{\sigma}_z$=0.5 mm and summed to create the pseudo-Green function wake~\cite{ishibashi2023impedance} as shown in Fig.~\ref{fig:wakeSKBLER}.
The corresponding impedances shown in the same figure are calculated by CZT of the wake data. The decay of impedance data at high frequencies is due to the 0.5-mm Gaussian bunch chosen, which acts as a Gaussian window function of $e^{-k^2\hat{\sigma}_{z}^2/2}$. This is justified when the ratio of the nominal length of the bunch $\sigma_{z0}$ to the length of the driving bunch $\hat{\sigma}_z$ is very large (for our example, 9.2). Additionally, the bunch should not be strongly deformed or micro-bunched, so the high-frequency impedances are not sampled. The beam parameters used to solve the Haissinski equation are summarized in Table~\ref{tb:LER-parameters}. The scaling parameter at $N=10^{11}$ is $I=0.0364$ pC/V. The numerically obtained Haissinski solutions with bunch populations $N=(0.0275, 5.49, 11.0, 16.5)\times 10^{10}$ (corresponding to $I$=(0.0001, 0.02, 0.04, 0.06) pC/V) as shown in Fig.~\ref{fig:haiDensity}.
The lengthening of the bunch, the centroid shifts, and the density peak as a function of the normalized current are shown in Fig.~\ref{fig:sigz}.
\begin{figure}
\centering
\includegraphics[width=0.85\linewidth]{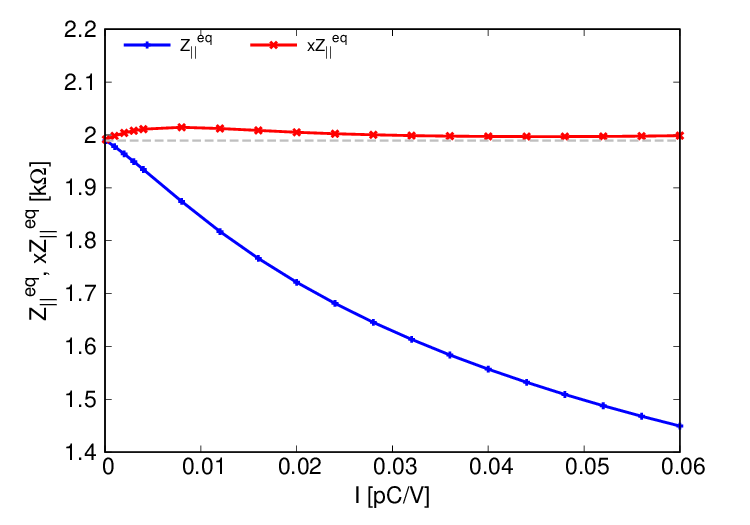}
\caption{Equivalent impedance as a function of normalized current calculated from simulated bunch lengths for SuperKEKB LER. The gray dashed line shows the equivalent impedance corresponding to Zottter's cubic equation.}
\label{fig:ZeffSKB}
\end{figure}
The equivalent impedances calculated from the quadratic and cubic scaling laws are plotted in Fig.~\ref{fig:ZeffSKB}. It is seen that inductive impedance largely dominates bunch lengthening in SuperKEKB LER, as can also be seen from the bunch profiles in Fig.~\ref{fig:haiDensity}.

\begin{figure}
\centering
\includegraphics[width=0.85\linewidth]{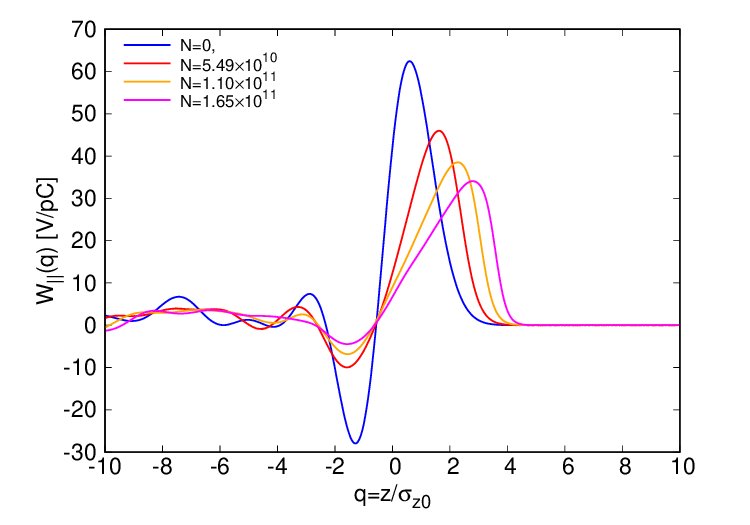}
\caption{Wake potentials calculated from Haissinski solutions of Fig.~\ref{fig:haiDensity} using Eq.~(\ref{eq:WakePotentialhai}), compared with the wake potential of a nominal bunch.}
\label{fig:WzHai}
\end{figure}

\begin{figure}
\centering
\includegraphics[width=0.85\linewidth]{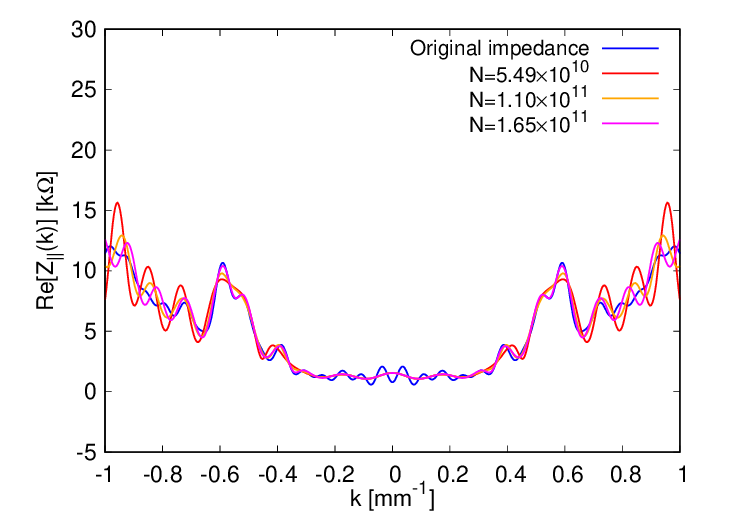}
\includegraphics[width=0.85\linewidth]{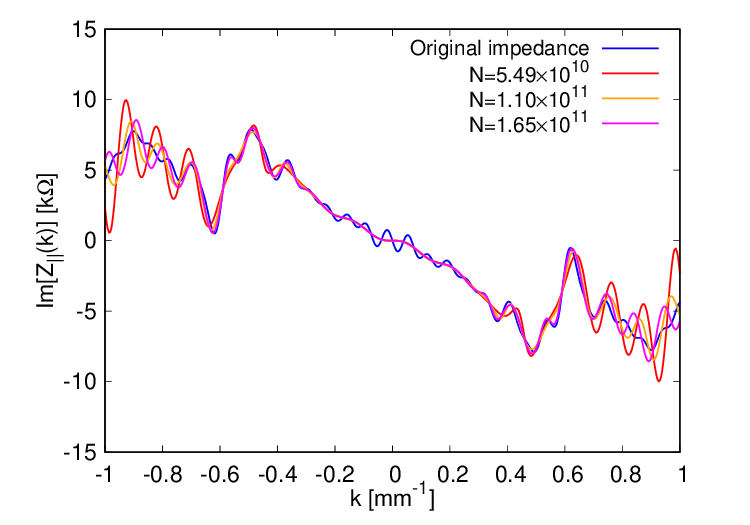}
\caption{Impedances calculated from Haissinski solutions of Fig.~\ref{fig:haiDensity} using Eq.~(\ref{eq:InvHaissinski1}), compared with the original impedance including CSR.}
\label{fig:impedanceSKBLERhai}
\end{figure}

The wake potentials and impedances extracted from the Haissinski solutions in different bunch populations are compared in Figs.~\ref{fig:WzHai} and~\ref{fig:impedanceSKBLERhai}. The extracted impedance does not contain the oscillatory impedance at low frequencies ($|k|\lesssim 200 \text{ m}^{-1}$ in Fig.~\ref{fig:impedanceSKBLERhai}). This is because the bunch does not experience long-range wake fields at $|z| \gg \sigma_{z0}$, as shown in Fig.~\ref{fig:wakeSKBLER}. Ideally, the bunch profiles of different currents should produce identical impedances at any frequency. However, the divergence of impedances at high frequencies can be seen in Fig.~\ref{fig:impedanceSKBLERhai}, demonstrating the impact of numerical noise on the simulated bunch profiles. For example, at a frequency of $k= 4/\sigma_{z0}\approx 870\text{ m}^{-1}$, the relative spectrum density is of the order of $\tilde{\lambda}_h(k) \approx e^{-8}\approx 3.3\times 10^{-4}$ if there is no obvious micro-bunching in the bunch profile. Such small amplitudes of spectrum density will amplify the noises in the spectrum of wake potential of Eq.~(\ref{eq:InvHaissinski3}) and consequently cause large errors in the impedance of Eq.~(\ref{eq:InvHaissinski1}). Furthermore, noise effects can also be significant when the deformations of bunch profiles are weak at low currents. This can be seen from Eq.~(\ref{eq:WakePotentialhai}): The sum within the square brackets approaches zero when $I\rightarrow 0$. Although the right-hand side should eventually converge, numerical noise could cause a divergence at $I=0$.

The divergence of extracted impedance at high frequencies also implies that the numerical noises in the bunch profile, but not the physical micro-bunching, have to be well controlled in Vlasov or macroparticle tracking simulations. Otherwise, the noise (corresponding to the high-frequency spectrum) may sample the high-frequency impedance and drive numerical instabilities. In other words, including physical high-frequency impedances as much as possible in impedance modeling will challenge simulation codes in their techniques of controlling numerical noise, such as a large number of macroparticles, fine meshing, smoothing algorithms, etc. From this point of view, calculating the impedance from the simulated bunch profile, as shown in Fig.~\ref{fig:impedanceSKBLERhai}, serves as a consistency check of the simulations.

Similar to KEKB LER, a substantial discrepancy persists between the measured bunch lengthening using a streak camera and the simulations in SuperKEKB LER~\cite{ishibashi2023impedance}. Although there is a discrepancy in measured and simulated bunch lengths, two experimental observations lend support to the reliability of the numerically constructed impedance model: 1) The simulated betatron tune shift using the impedance model (that is, longitudinal and transverse impedances are simultaneously included in simulations) agrees well with the measurements. 2) The simulated peak shift in the bunch profile using the impedance model is fairly in agreement with the measured synchronous phase shift using BPM signals. For 1), details are provided in~\cite{ishibashi2023impedance}. 2) is briefly discussed here, while the details of phase-shift measurement techniques will be presented in a separate paper.
\begin{figure}
\centering
\includegraphics[width=0.85\linewidth]{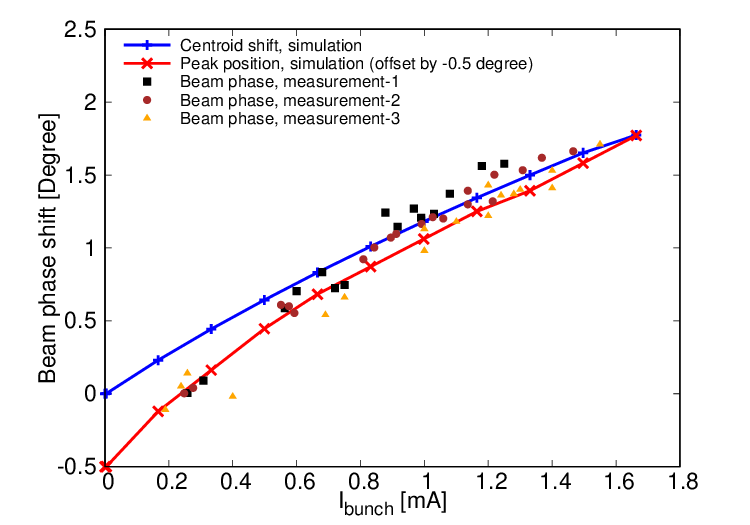}
\caption{Measured beam phase compared with simulations at SuperKEKB LER. Measurements 1 and 2 were performed in June 2021, and measurement 3 was performed in June 2020.}
\label{fig:BeamPhaseMeasurement}
\end{figure}
\begin{figure}
\centering
\includegraphics[width=0.85\linewidth]{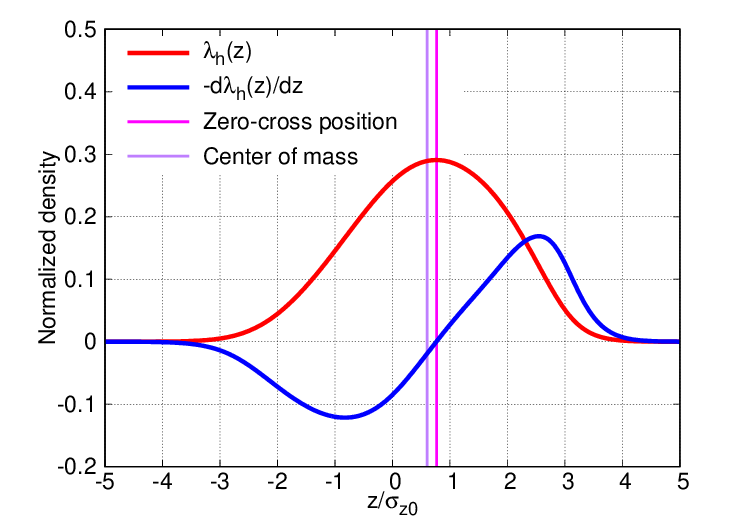}
\caption{A simulated bunch profile $\lambda_h(z)$ with potential-well distortion and its derivative $-\lambda_h'(z)$. The zero-crossing point of $-\lambda_h'(z)$ corresponds to the peak of $\lambda_h(z)$ (marked by the magenta line), but not the center of mass (marked by the purple line).}
\label{fig:ZeroCrossingMeasurement}
\end{figure}

There are various techniques to measure the synchronous phase in electron storage rings (for example, see~\cite{farias2001oscilloscope, ieiri2009beam, cheng2017precise}).
For SuperKEKB LER, a simple method utilizing an oscilloscope was used to monitor the beam signal and detect its zero-crossing point to measure the synchronous phase. The results are compared with the centroid shift and the peak shift using Vlasov simulations with the impedance model, as shown in Fig.~\ref{fig:BeamPhaseMeasurement}. The three measurements displayed good general reproducibility of the current-dependent beam phase, albeit with some data scattering, potentially arising from machine conditions or experimental setup specifics. As the measurements did not provide absolute beam phases, the peak positions in the simulations were adjusted by -0.5 degrees, revealing a relatively good alignment with the measurements. This suggests that the impedance model predicts the measured behavior well. However, a notable discrepancy in centroid shift was observed, particularly at lower bunch currents. Although we could simply adjust the simulated centroid shifts at low currents to match the measurements, disagreement at high bunch currents will indicate a significant lack of resistance in the impedance model. Intuitively speaking, if the BPM signal is roughly proportional to $d\lambda_h(z)/dz$, as seen from Fig.~10 of~\cite{ieiri2009beam}, measuring the zero-crossing point is essentially measuring the peak position of the bunch profile, but not the center of mass. The principle is illustrated in Fig.~\ref{fig:ZeroCrossingMeasurement}. However, this correlation requires validation through a detailed simulation of the BPM signal, involving accurate modeling of both the button electrodes of the BPM and the long coaxial cable connecting the BPM to the oscilloscope. This detailed investigation is planned for a future study at SuperKEKB.

\section{\label{sec:summary}Summary}
In summary, the theory of impedance effects through potential-well distortion is revisited in this paper. The Haissinski equation, a self-consistent solution of the VFP equation below the microwave instability threshold, has been used to derive some theories to describe the intensity-dependent behaviors of measurable quantities. The effective impedance used to describe bunch lengthening has simply been replaced by a term of equivalent impedance in a self-consistent manner. 

When comparing impedance calculations and beam measurements, it is crucial to ensure that the measured and simulated quantities match or closely approximate each other with an acceptable systematic bias. The theories derived from the Haissinski equation are useful to facilitate such comparisons. The experience of KEKB and SuperKEKB highlights the importance of thoroughly understanding the principles behind beam-phase measurement techniques before comparing them with numerical simulations.

Measurements using BPMs or streak cameras can be used to extract various aspects of machine impedances. In principle, streak camera measurements can simultaneously predict loss factors and equivalent impedance. In particular, if the noises in the measured profile of a single bunch are well suppressed, it is possible to extract frequency-dependent impedances, as we have demonstrated using numerically obtained Haissinski solutions. This will be very useful for validating impedance computations or for looking for missing impedance sources from constructed impedance models of electron storage rings.

\begin{acknowledgments}
The author D.Z. thanks many colleagues, including M. Blaskiewicz, A. Blednykh, Y.-C Chae, A. Chao, Y. Cai, L. Carver, K. Hirata, R. Lindberg, M. Migliorati, K. Ohmi, K. Oide, B. Podobedov, Y. Shobuda, and V. Smaluk, for inspiring discussions on various aspects of impedance issues in electron storage rings.
\end{acknowledgments}

\appendix

\section{\label{sec:capacitor}On purely capacitive wake function}

 The reader may note that the wake function corresponding to Eq.~(\ref{eq:capacitiveZ}) by Fourier transform is
\begin{equation}
    W_\parallel(z)=\frac{1}{2C}\text{Sign}(-z)
    \label{eq:capacitiveW1}
\end{equation}
with $\text{Sign}(z)$ the sign function. Equation~(\ref{eq:capacitiveW1}) is not a causal wake function and is thus different from the conventionally defined causal wake function of (for example, see page.311 of~\cite{chao2023handbook})
\begin{equation}
    W_\parallel(z)=\frac{1}{C}H(-z)
    \label{eq:capacitiveW2}
\end{equation}
with $H(z)$ the Heaviside step function. The impedance corresponding to Eq.~(\ref{eq:capacitiveW2}) via Fourier transform is
\begin{equation}
    Z_\parallel(k)=
    \frac{i}{kcC} + \frac{\pi \delta(k)}{cC}.
    \label{eq:capacitiveZ2}
\end{equation}
The impedance of Eq.~(\ref{eq:capacitiveZ2}) is singular at $k=0$. To remove such a singularity, an alternative way is to use the Laplace transform to calculate the corresponding impedance of a causal wake function (note that it is defined at $z<0$ in our notation), i.e.
\begin{equation}
    Z_{\parallel\mathcal{L}}(s)=
    \int_{-\infty}^0 W_\parallel(z) e^{sz}dz.
    \label{eq:laplaceZ}
\end{equation}
Applying Eq.~(\ref{eq:laplaceZ}) to Eq.~(\ref{eq:capacitiveW2}), we can obtain a Laplace impedance of
\begin{equation}
    Z_{\parallel\mathcal{L}}(k)=\frac{1}{sC}.
    \label{eq:laplaceZ2}
\end{equation}
This is a smooth function, but one needs to use inverse Laplace transformation when calculating its wake function or corresponding wake potentials of a bunch.


\bibliography{aps_hai}

\end{document}